\newif\iflongversion
\longversiontrue

\iflongversion
\documentclass[10pt,twoside]{article}
\usepackage{fullpage}
\usepackage{times}
\usepackage{doi}

\usepackage{titlesec}

\titleformat{\subsubsection}[runin]
{\normalfont\bfseries}{\thesubsection}{1em}{}

\usepackage{fancyhdr}
\usepackage{amsthm}

\newtheorem{theorem}{Theorem}
\newtheorem{lemma}[theorem]{Lemma}
\newtheorem{proposition}[theorem]{Proposition}

\theoremstyle{remark}

{\itshape}{}

\let\oldbibliography\thebibliography
\renewcommand{\thebibliography}[1]{%
  \oldbibliography{#1}%
  \setlength{\itemsep}{0pt}%
}

\else
\documentclass[runningheads,envcountsame]{llncs}

\spnewtheorem{counterexample}[example]{Counterexample}{\itshape}{}

\RequirePackage[bookmarks,unicode,colorlinks=true]{hyperref}%

\makeatletter
\RequirePackage[bookmarks,unicode,colorlinks=true]{hyperref}%
   \def\@citecolor{blue}%
   \def\@urlcolor{blue}%
   \def\@linkcolor{blue}%

\def\orcidID#1{\href{http://orcid.org/#1}{\protect\raisebox{-1.25pt}{\protect\includegraphics{orcid.eps}}}}
\makeatother

\fi

\usepackage{graphicx}

\usepackage{xcolor}
\usepackage{paralist}

\usepackage[caption=false]{subfig}

\usepackage{wrapfig}
\usepackage{blindtext}
\usepackage{etoolbox}
\BeforeBeginEnvironment{wrapfigure}{\setlength{\intextsep}{0pt}}

\usepackage{bm}
\usepackage{marvosym}

\usepackage[bbsets,Dfprime,setrelation]{math}
\usepackage{manalysis}

\usepackage{wasysym}
\usepackage[prefixflatinterpret,bracketmodalinterpret,fixformat,silentconst,sidenotecalculus,longseqcontext,seqinsist]{logic}
\usepackage[prefixflatinterpret,bracketmodalinterpret,fixformat,silentconst,differentialdL,simplenames]{dL}
\newcommand{\bebecomes}{\mathrel{::=}}
\newcommand{\alternative}{~|~}

\newcommand{\I}{\dLint[const=I,state=\nu]}

\usepackage{ upgreek }

\newcommand{\rfvar}{P}

\newcommand{\invvar}{I}

\usepackage{prettyref}
\newcommand{\rref}[2][]{\prettyref{#2}}
\newrefformat{sec}{Section\,\ref{#1}}
\newrefformat{subsec}{Section\,\ref{#1}}
\newrefformat{def}{Def.\,\ref{#1}}
\newrefformat{thm}{Theorem\,\ref{#1}}
\newrefformat{prop}{Proposition\,\ref{#1}}
\newrefformat{rem}{Remark\,\ref{#1}}
\newrefformat{lem}{Lemma\,\ref{#1}}
\newrefformat{cor}{Corollary\,\ref{#1}}
\newrefformat{ex}{Example\,\ref{#1}}
\newrefformat{cex}{Counterexample\,\ref{#1}}
\newrefformat{tab}{Table\,\ref{#1}}
\newrefformat{fig}{Fig.\,\ref{#1}}
\newrefformat{case}{case\,\ref{#1}}
\newrefformat{foot}{Footnote\,\ref{#1}}

\iflongversion
\newrefformat{app}{Appendix\,\ref{#1}}
\else
\newrefformat{app}{\cite{DBLP:journals/corr/abs-2203-01272}}
\fi

\newcommand{\folr}{FOL$_{\mathbb{R}}$\xspace}

\newsavebox{\Lightningval}%
\sbox{\Lightningval}{\mbox{\lightning}}

\newsavebox{\Rval}%
\sbox{\Rval}{$\scriptstyle\mathbb{R}$}
\relax

  \newdimen\linferenceRulehskipamount%
  \newdimen\lcalculuscollectionvskipamount%
\definecolor{vblue}{rgb}{.1,.15,.62}

\definecolor{vgreen}{rgb}{.1,.5,0}
\definecolor{vgray}{rgb}{.45,.45,.45}
\definecolor{vred}{rgb}{.7,0,0}

\newcommand{\redc}[1]{\textcolor{vred}{#1}}

\renewcommand*{\irrulename}[1]{\text{\textcolor{vgray}{\upshape#1}}}%
\renewcommand{\linferenceRuleNameSeparation}{\hspace{12pt}}
\renewcommand{\linferenceRuleSidenoteSeparation}{\quad}
\renewcommand{\linferenceRuleRefSeparation}{, }

\renewcommand*{\lie}[3][]
{\mathcal{L}_{#2}^{\ifthenelse{\equal{#1}{}}{}{^{\left(#1\right)}}}(#3)}
\renewcommand*{\lied}[3][]{\overset{\bm .}{#3}\ifthenelse{\equal{#1}{}}{}{{}^{(#1)}}}

\renewcommand{\Dostar}[1]{\ifthenelse{\equal{#1}{}}{(*)}{-(*)}}

\newcommand{\etermA}{e}
\newcommand{\etermB}{\tilde{e}}

\newcommand{\fvarA}{\phi}
\newcommand{\fvarB}{\psi}

\newcommand{\funcsym}{h}
\newcommand{\interpfunc}[2][\funcsym]{{#1}_{{\ll}#2{\gg}}}

\newcommand{\exswing}{\ensuremath{\alpha_s}}
\newcommand{\exswingbad}{\ensuremath{\hat{\alpha}_s}}
\newcommand{\exswingfml}{\ensuremath{\fvarA_s}\xspace}

\newcommand{\exball}{\ensuremath{\alpha_b}}
\newcommand{\exballfml}{\ensuremath{\fvarA_b}\xspace}

\newcommand{\exneuron}{\ensuremath{\alpha_n}}
\newcommand{\exneuronfml}{\ensuremath{\fvarA_n}\xspace}

\newcommand{\explane}{\ensuremath{\alpha_a}}

\newcommand{\invariant}{\textit{Inv}}

\usepackage{listings}
\usepackage{keylistings}
\usepackage[T1]{fontenc}
\usepackage[scaled=0.85]{beramono}

\definecolor{codegray}{rgb}{0.5,0.5,0.5}

\lstdefinelanguage{KeYmaeraX}{%
  keywords={},
  sensitive=true,
  basicstyle=\scriptsize\ttfamily\upshape,
  morecomment=[s]{/*}{*/},
  morestring=[b]",
  morestring=*[s][\color{gray}]{==\"}{\"},
  morestring=*[s][\color{gray}]{~=\"}{\"},
  morestring=[s][\color{gray}\it\underbar]{\#}{\#},
  deletestring=[b]',
  deletestring=[m]{"},
  showstringspaces=false,
  commentstyle=\color{vgreen},
  stringstyle=\color{vblue},
  identifierstyle=\upshape,
  numberstyle=\tiny\color{codegray},
  mathescape,
  breaklines=true,
  postbreak=\mbox{$\hookrightarrow$\space},
  numbers=left,
  frame=single,
  otherkeywords = {'R,'L},
  morekeywords = [2]{if,then,else,Real,Bool,HP,Functions,ProgramVariables,Problem,End,Definitions,ArchiveEntry,Tactic,@invariant,implicit},
  keywordstyle = [2]{\bf\color{black}},
  morekeywords = [3]{'R,'L},
  keywordstyle = [3]{\color{gray}},
  morekeywords = [4]{implyR,loop,id,QE,composeb,solve,assignb,choiceb,andR,testb,auto},
  keywordstyle = [4]{\color{purple}},
  escapeinside={/*@}{@*/}
}[comments,keywords,strings]

\usepackage{etoolbox} %

\begin{document}

\iflongversion
\title{Implicit Definitions with Differential Equations for \KeYmaeraX \\ (System Description)}

\author{
James Gallicchio \and
Yong Kiam Tan \and
Stefan Mitsch \and
Andr\'e Platzer
\thanks{
  Computer Science Department, Carnegie Mellon University, Pittsburgh, USA \newline
  James Gallicchio: jgallicc@andrew.cmu.edu ; Yong Kiam Tan, Stefan Mitsch, Andr\'e Platzer: {\{yongkiat,smitsch,aplatzer\}@cs.cmu.edu}
  }
}
\date{}
\else
\title{Implicit Definitions with Differential Equations for \KeYmaeraX}
\subtitle{(System Description)}
\author{
James Gallicchio\orcidID{0000-0002-0838-3240}
\and Yong Kiam Tan\orcidID{0000-0001-7033-2463}
\and Stefan Mitsch\orcidID{0000-0002-3194-9759}
\and Andr\'e Platzer\orcidID{0000-0001-7238-5710}
}%
\authorrunning{J. Gallicchio et al.}

\institute{
Computer Science Department, Carnegie Mellon University, Pittsburgh, USA\\
\email{jgallicc@andrew.cmu.edu} \\
\email{\{yongkiat,smitsch,aplatzer\}@cs.cmu.edu}
}
\fi

\maketitle              %
\begin{abstract}
Definition packages in theorem provers provide users with means of defining and organizing concepts of interest.
This system description presents a new definition package for the hybrid systems theorem prover \KeYmaeraX based on differential dynamic logic (\dL).
The package adds \KeYmaeraX support for user-defined smooth functions whose graphs can be implicitly characterized by \dL formulas.
Notably, this makes it possible to implicitly characterize functions, such as the exponential and trigonometric functions, as solutions of differential equations and then prove properties of those functions using \dL's differential equation reasoning principles.
Trustworthiness of the package is achieved by minimally extending \KeYmaeraX's soundness-critical kernel with a single axiom scheme that expands function occurrences with their implicit characterization.
Users are provided with a high-level interface for defining functions and non-soundness-critical tactics that automate low-level reasoning over implicit characterizations in hybrid system proofs.

\iflongversion
\else
\fi
\end{abstract}

\section{Introduction}
\label{sec:introduction}

\KeYmaeraX \cite{DBLP:conf/cade/FultonMQVP15} is a theorem prover implementing differential dynamic logic \dL \cite{DBLP:journals/jar/Platzer08,DBLP:conf/lics/Platzer12b,DBLP:journals/jar/Platzer17,Platzer18} for specifying and verifying properties of hybrid systems mixing discrete dynamics and differential equations.
Definitions enable users to express complex theorem statements in concise terms, e.g., by modularizing hybrid system models and their proofs~\cite{DBLP:journals/corr/abs-2108-02965}.
Prior to this work, \KeYmaeraX had only one mechanism for definition, namely, non-recursive abbreviations via uniform substitution~\cite{DBLP:journals/corr/abs-2108-02965,DBLP:journals/jar/Platzer17}.
This restriction meant that common and useful functions, e.g., the trigonometric and exponential functions, could not be directly used in \KeYmaeraX, even though they can be uniquely characterized by \dL formulas~\cite{DBLP:journals/jar/Platzer08}.

This system description introduces a new \KeYmaeraX definitional mechanism where functions are \emph{implicitly defined} in \dL as solutions of ordinary differential equations (ODEs).
Although definition packages are available in most general-purpose proof assistants, our package is novel in tackling the question of how best to support user-defined functions in the \emph{domain-specific} setting for hybrid systems.
In contrast to tools with builtin support for \emph{some} fixed subsets of special functions~\cite{DBLP:journals/jar/AkbarpourP10,DBLP:conf/cade/GaoKC13,DBLP:journals/tecs/RatschanS07}; or higher-order logics that can work with functions via their infinitary series expansions~\cite{DBLP:journals/mscs/BoldoLM16}, e.g., $\exp(t) \mnodefeq \sum_{i=0}^\infty \frac{t^i}{i!}$; our package strikes a balance between practicality and generality by allowing users to define and reason about \emph{any} function characterizable in \dL as the solution of an ODE (\rref{sec:background}), e.g., $\exp(t)$ solves the ODE $\D{e}=e$ with initial value $e(0)=1$.

Theoretically, implicit definitions strictly expand the class of ODE invariants amenable to \dL's complete ODE invariance proof principles~\cite{DBLP:journals/jacm/PlatzerT20}; such invariants play a key role in ODE safety proofs~\cite{Platzer18} (see~\rref{prop:expressivity}).
In practice, arithmetical identities and other specifications involving user-defined functions are proved by automatically unfolding their implicit ODE characterizations and re-using existing \KeYmaeraX support for ODE reasoning (\rref{sec:implementation}). %
The package is designed to provide seamless integration of implicit definitions in \KeYmaeraX and its usability is demonstrated on several hybrid system verification examples drawn from the literature that involve special functions (\rref{sec:examples}).

\iflongversion
All proofs and examples are in Appendix~\ref{app:proofs} and~\ref{app:examples}. The definitions package is part of \KeYmaeraX with a usage guide at:
\url{http://keymaeraX.org/keymaeraXfunc/}.
\else
All proofs are in the supplement~\rref{app:}. The definitions package is part of \KeYmaeraX with a usage guide at:
\url{http://keymaeraX.org/keymaeraXfunc/}.
\fi

\section{Interpreted Functions in Differential Dynamic Logic}
\label{sec:background}

This section briefly recalls differential dynamic logic (\dL)~\cite{DBLP:journals/jar/Platzer08,Platzer10,DBLP:journals/jar/Platzer17,Platzer18} and explains how its term language is extended to support implicit function definitions.

\subsubsection{Syntax.}
Terms $\etermA, \etermB$ and formulas $\fvarA,\fvarB$ in \dL are generated by the following grammar, with variable $x$, rational constant $c$, $k$-ary function symbols $\funcsym$ (for any $k \in \naturals$), comparison operator $\sim {\in}~\{=,\neq,\geq,>,\leq,<\}$, and hybrid program $\alpha$:
{\small\begin{align}
  \etermA,\etermB &\bebecomes x \alternative c \alternative \etermA + \etermB \alternative \etermA \cdot \etermB \alternative \funcsym(\etermA_1,\dots,\etermA_k)  \label{eq:terms} \\
  \fvarA,\fvarB &\bebecomes \etermA \sim \etermB \alternative \fvarA \land \fvarB \alternative \fvarA \lor \fvarB \alternative \lnot{\fvarA} \alternative \lforall{x}{\fvarA} \alternative \lexists{x}{\fvarA} \alternative \dbox{\alpha}{\fvarA} \alternative\ddiamond{\alpha}{\fvarA} \label{eq:fmls}
\end{align}}%

The terms and formulas above extend the first-order language of real arithmetic (\folr) with the box ($\dbox{\alpha}{\fvarA}$) and diamond ($\ddiamond{\alpha}{\fvarA}$) modality formulas which express that \emph{all} or \emph{some} runs of hybrid program $\alpha$ satisfy postcondition $\fvarA$, respectively.
\rref{tab:hybrid-programs} gives an intuitive overview of \dL's hybrid programs language for modeling systems featuring discrete and continuous dynamics and their interactions thereof.
In \dL's uniform substitution calculus, function symbols $\funcsym$ are \emph{uninterpreted}, i.e., they semantically correspond to an arbitrary (smooth) function.
Such uninterpreted function symbols (along with uninterpreted predicate and program symbols) are crucially used to give a parsimonious axiomatization of \dL based on uniform substitution~\cite{DBLP:journals/jar/Platzer17} which, in turn, enables a trustworthy microkernel implementation of the logic in the theorem prover \KeYmaeraX~\cite{DBLP:conf/cade/FultonMQVP15,DBLP:series/lncs/MitschP20}.

\begin{table}[t]
\centering
\caption{Syntax and informal semantics of hybrid programs}
\label{tab:hybrid-programs}
\begin{tabular}{ll}
\hline
Program                & Behavior \\
\hline%
$\ptest{\fvarA}$         & Stay in the current state if $\fvarA$ is true, otherwise abort and discard run.\\
$\pumod{x}{\etermA}$   & Store the value of term $\etermA$ in variable $x$.\\
$\prandom{x}$          & Store an arbitrary real value in variable $x$.\\
$\pevolvein{\D{x}=\genDE{x}}{\ivr}$ & Continuously follow ODE $\D{x} = \genDE{x}$ in domain $\ivr$ for any duration ${\geq}0$. \\
$\pifs{\fvarA}{\alpha}$         & Run program $\alpha$ if $\fvarA$ is true, otherwise skip. Definable by \(\pchoice{\ptest{\fvarA};\alpha}{\ptest{\lnot\fvarA}}\).\\
$\alpha;\beta$         & Run program $\alpha$, then run program $\beta$ in any resulting state(s).\\
$\pchoice{\alpha}{\beta}$      & Nondeterministically run either program $\alpha$ or program $\beta$.\\
$\prepeat{\alpha}$             & Nondeterministically repeat program $\alpha$ for $n$ iterations, for any $n\in\mathbb{N}$.\\
\hline
$\{ \alpha \}$         & For readability, braces are used to group and delimit hybrid programs.\\
\hline
\end{tabular}
\end{table}

\begin{figure}[t!]
  \begin{minipage}[T]{0.25\textwidth}
  \includegraphics[width=\textwidth,clip,trim=0 0 0 0]{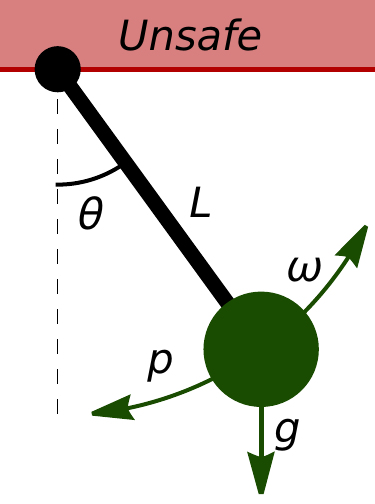}
  \end{minipage}\quad%
  \iflongversion\qquad\qquad\qquad\begin{minipage}[T]{0.5\textwidth}
  \else\begin{minipage}[T]{0.72\textwidth}\fi
  \small
  \textbf{Hybrid program model (auxiliary variables $\redc{s,c}$):}
  {\small\begin{flalign*}
  \exswingbad &\mnodefequiv \prepeat{\left(\begin{aligned}
  &\redc{\prandom{s}{}; \prandom{c}{}; \ptest{\fvarA_{\sin}(s,\theta)}; \ptest{\fvarA_{\cos}(c,\theta)};}\\
  &\prandom{p}{}; \keywordfont{if}\left(\frac{1}{2}(\omega-p)^2 < \frac{g}{L} \redc{c}\right)\, \{ \pumod{\omega}{\omega-p} \};\\
  &\{\pevolve{\D{\theta} = \omega, \D{\omega} = -\frac{g}{L}\redc{s} - k \omega, \redc{\D{s}=\omega c, \D{c}=-\omega s}}\}
  \end{aligned}\right)} &&
  \end{flalign*}}%
  \textbf{Hybrid program model (trigonometric functions):}
  {\small\begin{flalign*}
  \exswing &\mnodefequiv \prepeat{\left(\begin{aligned}
  &\prandom{p}{}; \keywordfont{if}\left(\frac{1}{2}(\omega-p)^2 < \frac{g}{L}\cos(\theta)\right)\, \{ \pumod{\omega}{\omega-p} \};\\
  &\{\pevolve{\D{\theta} = \omega, \D{\omega} = -\frac{g}{L}\sin(\theta) - k \omega}\}
  \end{aligned}\right)} &&
  \end{flalign*} }%
  \textbf{\dL safety specification:}
  {\small\begin{flalign*}
  \exswingfml &\mnodefequiv g > 0 \land L > 0 \land k > 0 \land\theta = 0 \land \omega = 0 \limply\dbox{\exswing}{\,\abs{\theta} < \frac{\pi}{2}}  &&
  \end{flalign*}}%
  \end{minipage}
  \caption{Running example of a swinging pendulum driven by an external force (left), its hybrid program models and \dL safety specification (right). Program $\exswing$ uses trigonometric functions directly, while program $\exswingbad$ uses variables $\redc{s,c}$ to implicitly track the values of $\sin(\theta)$ and $\cos(\theta)$, respectively (additions in \redc{red}). The implicit characterizations $\redc{\fvarA_{\sin}(s,\theta), \fvarA_{\cos}(c,\theta)}$ are defined in~\rref{eq:sindef},~\rref{eq:cosdef} and are not repeated here for brevity.}
  \label{fig:runningexample}
\end{figure}

\subsubsection{Running Example.}
Adequate modeling of hybrid systems often requires the use of \emph{interpreted} function symbols that denote specific functions of interest.
As a running example, consider the swinging pendulum shown in~\rref{fig:runningexample}.
The ODEs describing its continuous motion are $\pevolve{\D{\theta} = \omega, \D{\omega} = -\frac{g}{L}\sin(\theta) - k\omega}$, where $\theta$ is the swing angle, $\omega$ is the angular velocity, and $g, k, L$ are the gravitational constant, coefficient of friction, and length of the rigid rod suspending the pendulum, respectively.
The hybrid program $\exswing$ models an external force that repeatedly pushes the pendulum and changes its angular velocity by a nondeterministically chosen value $p$; the guard $\pifs{\dots}{}$ condition is designed to ensure that the push does not cause the pendulum to swing above the horizontal as specified by $\exswingfml$.
Importantly, the function symbols $\sin, \cos$ must denote the usual real trigonometric functions in $\exswing$.
Program $\exswingbad$ shows the same pendulum modeled in \dL \emph{without} the use of interpreted symbols, but instead using auxiliary variables $\redc{s, c}$.
Note that $\exswingbad$ is cumbersome and subtle to get right: the implicit characterizations $\redc{\fvarA_{\sin}(s,\theta), \fvarA_{\cos}(c,\theta)}$ from~\rref{eq:sindef},~\rref{eq:cosdef} are lengthy and the differential equations $\redc{\D{s}=\omega c, \D{c}=-\omega s}$ must be manually calculated and added to ensure that $\redc{s, c}$ correctly track the trigonometric functions as $\theta$ evolves continuously~\cite{Platzer10,DBLP:journals/jacm/PlatzerT20}.%

\subsubsection{Interpreted Functions.}
To enable extensible use of interpreted functions in \dL, the term grammar~\rref{eq:terms} is enriched with $k$-ary function symbols $\funcsym$ that carry an \emph{interpretation} annotation~\cite{DBLP:conf/lpar/BonichonDD07,DBLP:journals/corr/abs-1103-3322}, $\interpfunc[\funcsym]{\fvarA}$, where $\fvarA \mnodefequiv \fvarA(x_0,y_1,\dots,y_k)$ is a \dL formula with free variables in $x_0,y_1,\dots,y_k$ and no uninterpreted symbols.
Intuitively, $\fvarA$ is a formula that characterizes the graph of the intended interpretation for $\funcsym$, where $y_1,\dots,y_k$ are inputs to the function and $x_0$ is the output.
Since $\fvarA$ depends only on the values of its free variables, its formula semantics $\imodel{\I}{\fvarA}$ can be equivalently viewed as a subset of Euclidean space $\imodel{\I}{\fvarA} \subseteq \reals \times \reals^k$~\cite{DBLP:journals/jar/Platzer17,Platzer18}.
The \dL term semantics \(\ivaluation{\I}{\etermA}\)~\cite{DBLP:journals/jar/Platzer17,Platzer18} in a state $\iget[state]{\I}$ is extended with a case for terms \(\interpfunc{ \fvarA }(\etermA_1,\dots,\etermA_k)\) by evaluation of the smooth $C^\infty$ function characterized by \(\imodel{\I}{\fvarA}\):
{\begin{align*}
\ivaluation{\I}{\interpfunc{ \fvarA }(\etermA_1,\dots,\etermA_k)} =
\begin{cases}
\hat{\funcsym}(\ivaluation{\I}{\etermA_1},\dots,\ivaluation{\I}{\etermA_k}) & \text{if}~\imodel{\I}{\fvarA}~\text{graph of smooth $\hat{\funcsym} {:} \reals^k {\to} \reals$}\\
0 & \text{otherwise}
\end{cases}
\end{align*}}%

This semantics says that, if the relation $\imodel{\I}{\fvarA}  \subseteq \reals \times \reals^k$ is the graph of some smooth $C^\infty$ function $\hat{\funcsym} : \reals^k \to \reals$, then the annotated syntactic symbol $\interpfunc{ \fvarA }$ is interpreted semantically as $\hat{\funcsym}$. Note that the graph relation uniquely defines $\hat{\funcsym}$ (if it exists).
Otherwise, $\interpfunc{ \fvarA }$ is interpreted as the constant zero function which ensures that the term semantics remain well-defined for all terms.
An alternative is to leave the semantics of some terms (possibly) undefined, but this would require more extensive changes to the semantics of \dL and extra case distinctions during proofs~\cite{DBLP:conf/cade/BohrerFP19}.

\subsubsection{Axiomatics and Differentially-Defined Functions.}

To support reasoning for implicit definitions, annotated interpretations are reified to characterization axioms for expanding interpreted functions in the following lemma.

\begin{lemma}[Function interpretation]
\label{lem:fi}
The~\irref{FI} axiom (below) for \dL is sound where $\funcsym$ is a $k$-ary function symbol and the formula semantics \(\imodel{\I}{\fvarA}\) is the graph of a smooth $C^\infty$ function $\hat{\funcsym} : \reals^k \to \reals$.

\noindent
\[
\cinferenceRule[FI|FI]{function interpretation}
{\linferenceRule[equiv]
  {\fvarA(\etermA_0,\etermA_1,\dots,\etermA_k)}
  {\etermA_0 = \interpfunc[\funcsym]{\fvarA}(\etermA_1,\dots,\etermA_k)}
}
{}%
\]
\end{lemma}

Axiom~\irref{FI} enables reasoning for terms $\interpfunc[\funcsym]{\fvarA}(\etermA_1,\dots,\etermA_k)$ through their implicit interpretation $\fvarA$, but~\rref{lem:fi} does not directly yield an implementation because it has a soundness-critical side condition that interpretation $\fvarA$ characterizes the graph of a smooth $C^\infty$ function.
It is possible to syntactically characterize this side condition~\cite{DBLP:conf/cade/BohrerFP19}, e.g., the formula $\lforall{y_1,\dots,y_k}{\lexists{x_0}{\fvarA(x_0,y_1,\dots,y_k)}}$ expresses that the graph represented by $\fvarA$ has at least one output value $x_0$ for each input value $y_1,\dots,y_k$, but this burdens users with the task of proving this side condition in \dL before working with their desired function.
The \KeYmaeraX definition package opts for a middle ground between generality and ease-of-use by implementing~\irref{FI} for univariate, \emph{differentially-defined} functions, i.e., the interpretation $\fvarA$ has the following shape, where $x=(x_0,x_1,\dots,x_n)$ abbreviates a vector of variables, there is one input $t = y_1$, and $X = (X_0,X_1,\dots,X_n)$, $T$ are \dL terms that do not mention any free variables, e.g., are rational constants, which have constant value in any \dL state:
{\small\begin{align}
\fvarA(x_0,t) \mnodefequiv \ddiamond{ \prandom{x_1,\dots,x_n};\left\{\begin{array}{l}
\pevolve{\D{x}=-\genDE{x,t},\D{t}=-1}~\cup \\
\pevolve{\D{x}=\genDE{x,t},\D{t}=1}
\end{array}\right\}}{\left(\begin{aligned}
x&=X~\land \\
t&=T
\end{aligned} \right)}
\label{eq:ddef}
\end{align}}%

Formula~\rref{eq:ddef} says from point $x_0$, there exists a choice of the remaining coordinates $x_1,\dots,x_n$ such that it is possible to follow the defining ODE either forward \(\D{x}=\genDE{x,t},\D{t}=1\) or backward \(\D{x}=-\genDE{x,t},\D{t}=-1\) in time to reach the initial values $x=X$ at time $t=T$.
In other words, the implicitly defined function $\interpfunc[\funcsym]{\fvarA(x_0,t)}$ is the $x_0$-coordinate projected solution of the ODE starting from initial values $X$ at initial time $T$.
For example, the trigonometric functions used in~\rref{fig:runningexample} are differentially-definable as respective projections:
{\small\begin{align}
\fvarA_{\sin}(s,t) &\mnodefequiv \ddiamond{ \prandom{c};\left\{\begin{array}{l}
\pevolve{\D{s}=-c,\D{c}=\;\;\;s,\D{t}=-1}~\cup \\
\pevolve{\D{s}=\;\;\;c,\D{c}=-s,\D{t}=\;\,\,\,1}
\end{array}\right\}}{\left(\begin{aligned}
s&=0\land c=1~\land \\
t&=0
\end{aligned} \right)} \label{eq:sindef} \\
\fvarA_{\cos}(c,t) &\mnodefequiv \ddiamond{ \prandom{s};\left\{\begin{array}{l}
\pevolve{\D{s}=-c,\D{c}=\;\;\;s,\D{t}=-1}~\cup \\
\pevolve{\D{s}=\;\;\;c,\D{c}=-s,\D{t}=\;\,\,\,1}
\end{array}\right\}}{\left(\begin{aligned}
s&=0\land c=1~\land \\
t&=0
\end{aligned} \right)} \label{eq:cosdef}
\end{align}}%

By Picard-Lindel\"of~\cite[Thm.\,2.2]{Platzer18}, the ODE \( \D{x}=\genDE{x,t} \) has a unique solution $\Phi : (a,b) \to \reals^{n+1}$ on an open interval $(a,b)$ for some $-\infty \leq a < b \leq \infty$.
Moreover, $\Phi(t)$ is $C^\infty$ smooth in $t$ because the ODE right-hand sides are \dL terms with smooth interpretations~\cite{DBLP:journals/jar/Platzer17}.
Therefore, the side condition for~\rref{lem:fi} reduces to showing that $\Phi$ exists globally, i.e., it is defined on $t \in (-\infty,\infty)$.

\begin{lemma}[Smooth interpretation]
\label{lem:si}
If formula $\lexists{x_0}{\fvarA(x_0,t)}$ is valid, $\fvarA(x_0,t)$ from~\rref{eq:ddef} characterizes a smooth $C^\infty$ function and axiom~\irref{FI} is sound for $\fvarA(x_0,t)$.
\end{lemma}

\rref{lem:si} enables an implementation of axiom~\irref{FI} in \KeYmaeraX that combines a syntactic check (the interpretation has the shape of formula~\rref{eq:ddef}) and a side condition check (requiring users to prove existence for their interpretations).

The addition of differentially-defined functions to \dL strictly increases the deductive power of ODE invariants, a key tool in deductive ODE safety reasoning~\cite{Platzer18}.
Intuitively, the added functions allow direct, syntactic descriptions of invariants, e.g., the exponential or trigonometric functions, that have effective invariance proofs using \dL's complete ODE invariance reasoning principles~\cite{DBLP:journals/jacm/PlatzerT20}.

\begin{proposition}[Invariant expressivity]
\label{prop:expressivity}
There are valid polynomial \dL differential equation safety properties which are provable using differentially-defined function invariants but are not provable using polynomial invariants.
\end{proposition}

\section{\KeYmaeraX Implementation}
\label{sec:implementation}

The implicit definition package adds interpretation annotations and axiom~\irref{FI} based on~\rref{lem:si} in ${\approx}170$ lines of code extensions to \KeYmaeraX's soundness-critical core~\cite{DBLP:conf/cade/FultonMQVP15,DBLP:series/lncs/MitschP20}.
This section focuses on non-soundness-critical usability features provided by the package that build on those core changes.

\subsection{Core-Adjacent Changes}

\KeYmaeraX has a browser-based user interface with concrete, ASCII-based \dL syntax~\cite{DBLP:journals/corr/abs-2108-02965}.
The package extends \KeYmaeraX's parsers and pretty printers with support for interpretation annotations \texttt{h<<...>>(...)} and users can simultaneously define a family of functions as respective coordinate projections of the solution of an $n$-dimensional ODE (given initial conditions) with sugared syntax:
\[ \verb|implicit Real h1(Real t), ..., hn(Real t) = {{initcond};{ODE}}| \]

For example, the implicit definitions~\rref{eq:sindef},~\rref{eq:cosdef} can be written with the following sugared syntax; \KeYmaeraX automatically inserts the associated interpretation annotations for the trigonometric function symbols, see
\iflongversion
~\rref{app:examples} for a \KeYmaeraX snippet of formula \exswingfml from~\rref{fig:runningexample} using this sugared definition.
\else the supplement~\rref{app:examples} for a \KeYmaeraX snippet of formula \exswingfml from~\rref{fig:runningexample} using this sugared definition.
\fi
\iflongversion
\[ \verb|implicit Real sin(Real t), cos(Real t) = {{sin:=0; cos:=1;}; {sin'=cos, cos'=-sin}}|\]
\else
\begin{verbatim}
implicit Real sin(Real t), cos(Real t) =
    {{sin:=0; cos:=1;}; {sin'=cos, cos'=-sin}}
\end{verbatim}
\fi

In fact, the functions $\sin, \cos, \exp$ are so ubiquitous in hybrid system models that the package builds their definitions in automatically without requiring users to write them explicitly. %
In addition, although arithmetic involving those functions is undecidable~\cite{Goedel_1931,DBLP:journals/jsyml/Richardson68}, \KeYmaeraX can export those functions whenever its external arithmetic tools have partial arithmetic support for those functions.

\subsection{Intermediate and User-Level Proof Automation}
\label{subsec:impllemma}

The package automatically proves three important lemmas about user-defined functions that can be transparently re-used in all subsequent proofs:
\begin{enumerate}
\item It proves the side condition of axiom~\irref{FI} using \KeYmaeraX's automation for proving sufficient duration existence of solutions for ODEs~\cite{DBLP:journals/fac/TanP21} which automatically shows global existence of solutions for all affine ODEs and some univariate nonlinear ODEs.
As an example of the latter, the hyperbolic $\tanh$ function is differentially-defined as the solution of ODE $\D{x}=1-x^2$ with initial value $x=0$ at $t=0$ whose global existence is proved automatically.

\item \label{enum:init} It proves that the functions have initial values as specified by their interpretation, e.g., $\sin(0)=0$, $\cos(0)=1$, and $\tanh(0)=0$.

\item It proves the \emph{differential axiom}~\cite{DBLP:journals/jar/Platzer17} for each function that is used to enable syntactic derivative calculations in \dL, e.g., the differential axioms for $\sin,\cos$ are $\der{\sin(e)} = \cos(e)\der{e}$ and $\der{\cos(e)} = -\sin(e)\der{e}$, respectively.
Briefly, these axioms are automatically derived in a correct-by-construction manner using \dL's syntactic version of the chain rule for differentials~\cite[Fig. 3]{DBLP:journals/jar/Platzer17}, so the rate of change of $\sin(e)$ is the rate of change of $\sin(\cdot)$ with respect to its argument $e$, multiplied by the rate of change of its argument $\der{e}$.
\end{enumerate}

\iflongversion
These lemmas enable the use of differentially-defined functions alongside all existing ODE automation in \KeYmaeraX~\cite{DBLP:journals/jacm/PlatzerT20,DBLP:journals/fac/TanP21}.
\else
These lemmas enable the use of differentially-defined functions with all existing ODE automation in \KeYmaeraX~\cite{DBLP:journals/jacm/PlatzerT20,DBLP:journals/fac/TanP21}.
\fi
In particular, since differentially-defined functions are univariate Noetherian functions, they admit complete ODE invariance reasoning principles in \dL~\cite{DBLP:journals/jacm/PlatzerT20} as implemented in \KeYmaeraX.

The package also adds specialized support for arithmetical reasoning over differential definitions to supplement external arithmetic tools in proofs.
First, it allows users to manually prove identities and bounds using \KeYmaeraX's ODE reasoning.
For example, the bound $\tanh(\lambda x)^2 < 1$ used in the example $\exneuron$ from~\rref{sec:examples} is proved by \emph{differential unfolding} as follows
\iflongversion
(see~\rref{app:examples}):
\else
(see supplement~\rref{app:}):
\fi
{\small\begin{align*}
\linfer[]{
  \lsequent{}{\tanh(0)^2 < 1} \quad \lsequent{\tanh(\lambda v)^2 {<} 1}{\dbox{\pchoice{\{\pevolvein{\D{v}=1}{v \leq x}\}}{\{\pevolvein{\D{v}=-1}{v \geq x}\}}}{\tanh(\lambda v)^2 {<} 1}}
  }
  {\lsequent{}{\tanh(\lambda x)^2 < 1}}
\end{align*}}%

This deduction step says that, to show the conclusion (below rule bar), it suffices to prove the premises (above rule bar), i.e., the bound is true at $v=0$ (left premise) and it is preserved as $v$ is evolved forward $\D{v}=1$ or backward $\D{v}=-1$ along the real line until it reaches $x$ (right premise).
The left premise is proved using the initial value lemma for $\tanh$ while the right premise is proved by ODE invariance reasoning with the differential axiom for $\tanh$~\cite{DBLP:journals/jacm/PlatzerT20}.

Second, the package uses \KeYmaeraX's uniform substitution mechanism~\cite{DBLP:journals/jar/Platzer17} to implement (untrusted) abstraction of functions with fresh variables when solving arithmetic subgoals, e.g., the following arithmetic bound for example $\exneuron$ is proved by abstraction after adding the bounds $\tanh(\lambda x)^2 < 1, \tanh(\lambda y)^2 < 1$.
{\small\begin{align*}
\textbf{Bound:} \quad &x(\tanh(\lambda x)-\tanh(\lambda y)) + y (\tanh(\lambda x)+\tanh(\lambda y)) \leq 2\sqrt{x^2+y^2} \\
\textbf{Abstracted:} \quad &t_x^2 < 1 \land t_y^2 <1 \limply x(t_x-t_y) + y (t_x+t_y) \leq 2\sqrt{x^2+y^2}
\end{align*}}%

\section{Examples}
\label{sec:examples}

The definition package enables users to work with differentially-defined functions in \KeYmaeraX, including modeling and expressing their design intuitions in proofs.
This section applies the package to verify various continuous and hybrid system examples from the literature featuring such functions.

\paragraph{Discretely driven pendulum.} The specification \exswingfml~from~\rref{fig:runningexample} contains a discrete loop whose safety property is proved by a loop invariant, i.e., a formula that is preserved by the discrete and continuous dynamics in each loop iteration~\cite{Platzer18}.
The key invariant is $\invariant \mnodefequiv \frac{g}{L}(1-\cos{\theta})+ \frac{1}{2}\omega^2 < \frac{g}{L}$, which expresses that the total energy of the system (sum of potential and kinetic energy on the LHS) is less than the energy needed to cross the horizontal (RHS).
The main steps are as follows (proofs for these steps are automated by \KeYmaeraX):
\begin{enumerate}
\item $\invariant \limply \dbox{\keywordfont{if}\left(\frac{1}{2}(\omega-p)^2 < \frac{g}{L}\cos(\theta)\right)\, \{ \pumod{\omega}{\omega-p} \}}{\invariant}$, which shows that the discrete guard only allows push $p$ if it preserves the energy invariant, and
\item $\invariant \limply \dbox{\{\pevolve{\D{\theta} = \omega, \D{\omega} = -\frac{g}{L}\sin(\theta) - k\omega}\}}{\invariant}$, which shows that $\invariant$ is an energy invariant of the pendulum's ODE.
\end{enumerate}

\paragraph{Neuron interaction.} The ODE $\exneuron$ models the interaction between a pair of neurons~\cite{MR1201326}; its specification $\exneuronfml$ nests \dL's diamond and box modalities to express that the system norm ($\sqrt{x^2+y^2}$) is asymptotically bounded by $2 \tau$.
{\small\begin{align*}
\exneuron &\mnodefequiv \D{x}=-\frac{x}{\tau} + \tanh(\lambda x) - \tanh(\lambda y),\D{y}=-\frac{y}{\tau}  + \tanh(\lambda x) + \tanh(\lambda y)\\
\exneuronfml &\mnodefequiv \tau > 0 \limply
  \lforall{\varepsilon {>} 0}{
    \ddiamond{\exneuron}{\dbox{\exneuron}{\,\sqrt{x^2 +y^2} \leq 2\tau + \varepsilon}}
}{}
\end{align*}}%

The verification of $\exneuronfml$ uses differentially-defined functions in concert with \KeYmaeraX's symbolic ODE safety and liveness reasoning~\cite{DBLP:journals/fac/TanP21}.
The proof uses a decaying exponential bound $\sqrt{x^2+y^2} \leq \exp(-\frac{t}{\tau})\sqrt{x_0^2+y_0^2} + 2\tau(1-\exp(-\frac{t}{\tau}))$, where the constants $x_0, y_0$ are symbolic initial values for $x,y$ at initial time $t=0$, respectively.
Notably, the arithmetic subgoals from this example are all proved using abstraction and differential unfolding (\rref{sec:implementation}) without relying on external arithmetic solver support for $\tanh$.

\paragraph{Longitudinal flight dynamics.}

\begin{wrapfigure}{r}{.24\columnwidth}
  \includegraphics[width=.22\columnwidth]{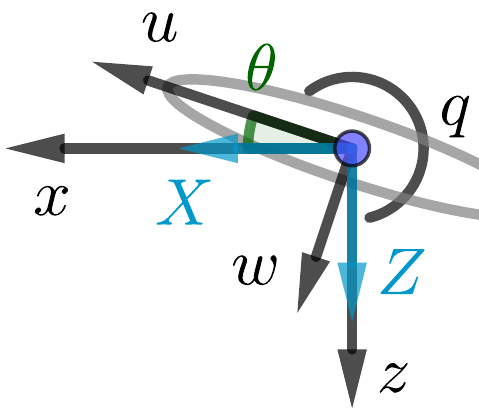}
\end{wrapfigure}

The differential equations $\explane$ below describe the 6th order longitudinal motion of an airplane while climbing or descending~\cite{DBLP:conf/tacas/GhorbalP14,Stengel2004}.
The airplane adjusts its pitch angle $\theta$ with pitch rate $q$, which determines its axial velocity $u$ and vertical velocity $w$, and, in turn, range $x$ and altitude $z$ (illustrated on the right).
The physical parameters are: gravity $g$, mass $m$, aerodynamic thrust and moment $M$ along the lateral axis, aerodynamic and thrust forces $X,Z$ along $x$ and $z$, respectively, and the moment of inertia $I_{yy}$, see~\cite[Section 6.2]{DBLP:conf/tacas/GhorbalP14}.
{\small\begin{align*}
  \explane \mnodefequiv \D{u}&=\frac{X}{m}-g \sin(\theta)-qw, &  \D{w}&=\frac{Z}{m}+g\cos(\theta)+qu, & \D{q}&=\frac{M}{I_{yy}}, \\
 \D{x}&=\cos(\theta)u+\sin(\theta)w, & \D{z}&=-\sin(\theta)u + \cos(\theta)w, & \D{\theta}&=q
\end{align*}}%

The verification of specification $J \limply \dibox{\explane}J$ shows that the safety envelope $J \equiv J_1 \land J_2 \land J_3$ is invariant along the flow of $\explane$ with algebraic invariants $J_i$:
{\small\begin{align*}
    J_1 &\equiv \frac{Mz}{I_{yy}} + g\theta + \left(\frac{X}{m}-qw\right)\cos(\theta) + \left(\frac{Z}{m}+qu\right)\sin(\theta)=0 \\
    J_2 &\equiv \frac{Mz}{I_{yy}} - \left(\frac{Z}{m}+qu\right)\cos(\theta) + \left(\frac{X}{m}-qw\right)\sin(\theta)=0 \quad J_3 \equiv -q^2 + \frac{2M\theta}{I_{yy}}=0
\end{align*}}%

\iflongversion
Additional examples are available in~\rref{app:examples}, including: a bouncing ball on a sinusoidal surface~\cite{DBLP:phd/ethos/Denman15,DBLP:conf/fm/0009ZZZ15} and a robot collision avoidance model~\cite{DBLP:journals/ijrr/MitschGVP17}.
\else
Additional examples are available in the supplement~\rref{app:examples}, including: a bouncing ball on a sinusoidal surface~\cite{DBLP:phd/ethos/Denman15,DBLP:conf/fm/0009ZZZ15} and a robot collision avoidance model~\cite{DBLP:journals/ijrr/MitschGVP17}.
\fi

\section{Conclusion}
\label{sec:conclusion}

This work presents a convenient mechanism for extending the \dL term language with differentially-defined functions, thereby furthering the class of real-world systems amenable to modeling and formalization in \KeYmaeraX.
Minimal soundness-critical changes are made to the \KeYmaeraX kernel, which maintains its trustworthiness while allowing the use of newly defined functions in concert with all existing \dL hybrid systems reasoning principles implemented in \KeYmaeraX.
Future work could formally verify these kernel changes by extending the existing formalization of \dL~\cite{DBLP:conf/cpp/BohrerRVVP17}.
Further integration of external arithmetic tools~\cite{DBLP:journals/jar/AkbarpourP10,DBLP:conf/cade/GaoKC13,DBLP:journals/tecs/RatschanS07} will also help to broaden the classes of arithmetic sub-problems that can be solved effectively in hybrid systems proofs.

\paragraph{Acknowledgments.}
We thank the anonymous reviewers for their helpful feedback on this paper.
This material is based upon work supported by the National Science Foundation under Grant No. CNS-1739629.
This research was sponsored by the AFOSR under grant number FA9550-16-1-0288.
\iflongversion
The views and conclusions contained in this document are those of the author and should not be interpreted as representing the official policies, either expressed or implied, of any sponsoring institution, the U.S. government or any other entity.
\fi

\iflongversion
\bibliographystyle{halpha}
\else
\bibliographystyle{splncs04}
\fi

\bibliography{paper}

\iflongversion
\newpage
\appendix

\section{Proofs}
\label{app:proofs}
This appendix presents proofs for the lemmas presented in~\rref{sec:background} that justify the soundness of axiom~\irref{FI} for differentially-defined functions.

\begin{proof}[Proof of~\rref{lem:fi}]
Let $\interpfunc{ \fvarA }$ be a $k$-ary function symbol with annotated interpretation $\fvarA \mnodefequiv \fvarA(x_0,x_1,\dots,x_k)$, where $\fvarA$ characterizes a smooth function, i.e., $\imodel{\I}{\fvarA} \subseteq \reals \times \reals^k$ is the graph of a smooth $C^\infty$ function $\hat{\funcsym} : \reals^k \to \reals$.
Soundness of axiom~\irref{FI} is shown by proving its equivalence true in an arbitrary \dL state $\iget[state]{\I}$.
The proof proceeds by calculation with the \dL semantics~\cite{DBLP:journals/jar/Platzer17,Platzer18} extended with term semantics for interpreted functions from~\rref{sec:background}.
\begin{align*}
\imodels{\I}{\etermA_0 = \interpfunc[\funcsym]{\fvarA}(\etermA_1,\dots,\etermA_k)}
&~\text{iff}~\ivaluation{\I}{\etermA_0} = \ivaluation{\I}{\interpfunc[\funcsym]{\fvarA}(\etermA_1,\dots,\etermA_k)}\\
&~\text{iff}~\ivaluation{\I}{\etermA_0} = \hat{\funcsym}(\ivaluation{\I}{\etermA_1},\dots,\ivaluation{\I}{\etermA_k})\\
&~\text{iff}~(\ivaluation{\I}{\etermA_0},\ivaluation{\I}{\etermA_1},\dots,\ivaluation{\I}{\etermA_k}) \in \imodel{\I}{\fvarA} \\
&~\text{iff}~\imodels{\I}{\fvarA(\etermA_0,\etermA_1,\dots,\etermA_k)}
\end{align*}

The final step uses that formula $\fvarA$ depends only on the values of its free variables so that its formula semantics $\imodel{\I}{\fvarA}$ can be equivalently viewed as a subset of Euclidean space $\imodel{\I}{\fvarA} \subseteq \reals \times \reals^k$~\cite{DBLP:journals/jar/Platzer17,Platzer18}.
\end{proof}

\begin{proof}[Proof of~\rref{lem:si}]
Let $\interpfunc{ \fvarA }$ be a unary function symbol with annotated interpretation $\fvarA \mnodefequiv \fvarA(x_0,t)$ according to the shape specified by formula~\rref{eq:ddef}.
To prove soundness of axiom~\irref{FI} for $\interpfunc{ \fvarA }$, by~\rref{lem:fi}, it suffices to show that $\imodel{\I}{\fvarA} \subseteq \reals \times \reals$ is the graph of a smooth $C^\infty$ function.

Let $\hat{X} = (\hat{X}_0,\hat{X}_1,\dots,\hat{X}_n) \in \reals^{n+1}, \hat{T} \in \reals$ denote the constant real value of terms $X = (X_0,X_1,\dots,X_n), T$, respectively.
By \dL formula semantics~\cite{DBLP:journals/jar/Platzer17,Platzer18}, the interpretation $\phi$ is true from state $(\hat{x}_0,\hat{t}) \in \reals^2$ iff there exists $\hat{x}_1,\dots,\hat{x}_n \in \reals^n$ such that either the solution of the forward ODE $\pevolve{\D{x}=\genDE{x,t},\D{t}=1}$ or the solution of the corresponding backward ODE $\pevolve{\D{x}=-\genDE{x,t},\D{t}=-1}$ reaches the initial state $(\hat{X},\hat{T})$.
Since solutions of the forward ODE are time-reversed solutions of the backward ODE (and vice-versa)~\cite{DBLP:journals/jacm/PlatzerT20} and variable $t$ tracks the forward progression of time with $\D{t}=1$ (or backward with $\D{t}=-1$), the interpretation $\phi$ is true in state $(\hat{x}_0,\hat{t})$ iff the solution of the nonautonomous ODE $\pevolve{\D{x}=\genDE{x,t}}$ from initial state $\hat{X}$ and time $\hat{T}$ reaches time $\hat{t}$ with value $\hat{x}_0$ for its $x_0$-coordinate (and there exist real values $\hat{x}_1,\dots,\hat{x}_n$ for the remaining coordinates).

By the Picard-Lindel\"of theorem~\cite[Thm.\,2.2]{Platzer18}, for any initial state $\hat{X}$ and initial time $\hat{T}$, the nonautonomous ODE \( \D{x}=\genDE{x,t} \) has a unique solution $\Phi(t) : (a,b) \to \reals^{n+1}$ on an open time interval $t \in (a,b)$ for some $-\infty \leq a < \hat{T} < b \leq \infty$.
In particular, formula $\phi$ is true in state $(\hat{x}_0,\hat{t})$ iff $\hat{t} \in (a,b)$ and the $x_0$-coordinate of $\Phi(\hat{t})$ is $\hat{x}_0$.
Moreover, $\Phi(t)$ is $C^\infty$ smooth in its argument $t$ because the ODE right-hand side $\genDE{x,t}$ are \dL terms with smooth interpretations~\cite{DBLP:journals/jar/Platzer17}.
Therefore, $\imodel{\I}{\fvarA}$ is the graph of the smooth function $\Phi_{x_0} : (a,b) \to \reals$ which projects the $x_0$-coordinate of solution $\Phi(t)$ at time $t \in (a,b)$.
Finally, by assumption, formula $\lexists{x_0}{\fvarA(x_0,t)}$ is valid, i.e., true for all values of free variable $t$. Thus, $\Phi_{x_0}(t)$ is defined for all $t \in (-\infty,\infty)$.
\end{proof}

\begin{proof}[Proof of~\rref{prop:expressivity}]
The \dL ODE safety property $\lsequent{\Gamma}{\dbox{\pevolvein{\D{x}=\genDE{x}}{\ivr}}{\rfvar}}$ expresses that, for all initial states satisfying assumptions $\Gamma$, all solutions of the ODE $\pevolve{\D{x}=\genDE{x}}$ from those states within domain $\ivr$ stay in the safe set characterized by formula $\rfvar$.
For the purposes of this proof, all formulas are assumed to only mention propositional connectives and (in)equalities over \dL terms, i.e., they \emph{neither} contain the first-order quantifiers nor \dL's dynamic modalities.
The safety property is called \emph{polynomial} if formulas $\Gamma, \ivr, \rfvar$ and ODE $\D{x}=\genDE{x}$ mention only polynomial terms, i.e., grammar~\rref{eq:terms} without additional function symbols.
The key technique for proving ODE safety properties is to find a suitable \emph{invariant} $\invvar$ of the ODE~\cite{Platzer18} such that \begin{inparaenum}[\it i)]
 \item it contains the initial states $\lsequent{\Gamma}{\invvar}$,
 \item it is safe $\lsequent{\invvar}{\rfvar}$, and
 \item solutions of the ODEs cannot escape it $\lsequent{\invvar}{\dbox{\pevolvein{\D{x}=\genDE{x}}{\ivr}}{\invvar}}$.
\end{inparaenum}
Formally, formula $\invvar$ is chosen such that all three premises of the following derived \dL proof rule are provable.
{%
\begin{sequentdeduction}[array]
\linfer[]{
  \lsequent{\Gamma}{\invvar} !
  \lsequent{\invvar}{\dbox{\pevolvein{\D{x}=\genDE{x}}{\ivr}}{\invvar}} ! \lsequent{\invvar}{\rfvar}
}
  {\lsequent{\Gamma}{\dbox{\pevolvein{\D{x}=\genDE{x}}{\ivr}}{\rfvar}}}
\end{sequentdeduction}
}%

The formula $\invvar$ is a polynomial invariant iff it only mentions polynomial terms.
Notably, \dL has \emph{complete} ODE invariance reasoning principles~\cite{DBLP:journals/jacm/PlatzerT20}, i.e., premise $\lsequent{\invvar}{\dbox{\pevolvein{\D{x}=\genDE{x}}{\ivr}}{\invvar}}$ is provably equivalent to an arithmetical formula in \dL; this completeness result holds not only for polynomial invariants but also more general term language extensions of \dL, including the Noetherian functions~\cite{DBLP:journals/jacm/PlatzerT20}, of which differentially-defined functions are a special case.
Consider the following polynomial ODE safety property:
\begin{align}
\underbrace{t=0 \land x \leq 1}_{\Gamma} \limply \dbox{\pevolve{\D{x}=x(2t-1),\D{t}=1}}{\underbrace{(t=1\limply x \leq 1)}_{\rfvar}} \label{eq:safetyprop}
\end{align}

The ODE $\D{x}=x(2t-1),\D{t}=1$ has an explicit solution with $x(\tau) = x_0\exp{((t_0 + \tau)^2-(t_0 + \tau))}, t(\tau) = t_0 + \tau$ for all times $\tau$ and initial values $x(0)=x_0, t(0)=t_0$.
The ODE safety property~\rref{eq:safetyprop} is provable using invariant $\invvar \mnodefequiv x \leq \exp{(t^2-t)}$, where $\exp(\cdot)$ is the differentially-definable real exponential function.

Suppose for contradiction that there is a polynomial invariant $\hat{\invvar}$ that proves property~\rref{eq:safetyprop}.
Thus, premises $\lsequent{\Gamma}{\hat{\invvar}}$, $\lsequent{\hat{\invvar}}{\rfvar}$, and $\lsequent{\hat{\invvar}}{\dbox{\pevolvein{\D{x}=\genDE{x}}{\ivr}}{\hat{\invvar}}}$ are valid.
This implies, semantically, that the set characterized by $\hat{\invvar}$ contains the forward trajectories from all initial points with $x(0) \leq 1$, i.e., all points satisfying formula $t \geq 0 \land x \leq \exp{(t^2-t)}$.
Moreover, all points with $0 \leq t \leq 1$ that satisfy formula $\hat{\invvar}$ must also satisfy formula $x \leq \exp{(t^2-t)}$.
To see this, suppose $x_0 > \exp{(t_0^2-t_0)}$, then from the explicit solution, when $t(\tau)=1$ and $\tau = 1-t_0$, $x(\tau) = x_0\exp{((t_0 + (1-t_0))^2-(t_0 + (1-t_0)))} = x_0 > \exp{(t(\tau)^2-t(\tau))}$, which violates the safety condition.
Thus, the set characterized by formula $0 \leq t \leq 1 \land \hat{\invvar}$ which restricts $\hat{\invvar}$ to the interval $0 \leq t \leq 1$ is equivalently characterized by $0 \leq t \leq 1 \land x \leq \exp{(t^2-t)}$.
However, this set is not semialgebraic~\cite[Definition 2.1.4]{Bochnak1998} so $\hat{\invvar}$ cannot be a polynomial invariant, contradiction.
\end{proof}

\section{Extended Examples}
\label{app:examples}

This appendix provides additional details for the examples which were elided in~\rref{sec:examples}.

\paragraph{Discretely driven pendulum (\KeYmaeraX model).} The formula \exswingfml from~\rref{fig:runningexample} is shown in the following \KeYmaeraX model snippet using a sugared definition which automatically inserts interpretation annotations for the trigonometric functions, following~\rref{eq:sindef},~\rref{eq:cosdef}.
The proof of this model is described in~\rref{sec:examples}.
\begin{lstlisting}[language=KeYmaeraX,numbers=none]
Definitions
  Real g, L, k;        /* gravity, length of rod, coeff. friction against angular velocity  */
  implicit Real sin(Real t), cos(Real t) '= {{sin:=0; cos:=1;}; {sin'=cos, cos'=-sin}};
  /* Desugared below (the above ODEs for sin, cos are omitted with ... for brevity) *
   * sin ~> sin<< <{cos:=*;sin:=._0;t:=._1;}{{...}++{...}}>(sin=0&cos=1&t=0) >>     *
   * cos ~> cos<< <{sin:=*;cos:=._0;t:=._1;}{{...}++{...}}>(sin=0&cos=1&t=0) >>     */
End.

ProgramVariables
  Real w, theta, push; /* angular velocity, displacement angle, extra push */
End.

Problem
  g > 0 & L > 0 & k > 0 &       /* Physical constants */
  theta = 0 & w = 0 ->         /* Start swing at rest */
  [{  /* Discrete push allowed if it is safe to do so */
    { push :=*; if (1/2*(w-push)^2 < g/L * cos(theta)) {w := w-push;} }
    { theta' = w, w' = -g/L * sin(theta) - k*w } /* Continuous dynamics */
  }*] (-pi()/2 < theta & theta < pi()/2)         /* Swing never crosses horizontal */
End.

\end{lstlisting}

\paragraph{Neuron interaction (differential unfolding).} The general \emph{differential unfolding} proof rule derived in \KeYmaeraX is as follows:
{\small\begin{align*}
\linfer[]{
  \lsequent{\Gamma}{\rfvar(v_0)} \quad \lsequent{\rfvar(v)}{\dbox{\pchoice{\{\pevolvein{\D{v}=1}{v \leq x}\}}{\{\pevolvein{\D{v}=-1}{v \geq x}\}}}{\rfvar(v)}}
  }
  {\lsequent{\Gamma}{\rfvar(x)}}
\end{align*}}%

This proof rule says that, to show the conclusion $\rfvar(x)$ (below rule bar), it suffices to prove the premises (above rule bar), i.e., the property $\rfvar(v_0)$ is true at an initial value $v=v_0$ (left premise) and $\rfvar(v)$ is preserved as $v$ is evolved forward $\D{v}=1$ or backward $\D{v}=-1$ along the real line until it reaches $x$ (right premise).
The proof rule is useful when formula $\rfvar$ mentions differentially-defined functions because it enables reasoning for those functions using properties of their implicit differential equations.
For example, the bound $\rfvar(x) \mnodefequiv \tanh(\lambda x)^2 < 1$ used in example $\exneuron$ from~\rref{sec:examples} is proved by differential unfolding on $x$ with $v_0 \mnodefeq 0$:
{\small\begin{align*}
\linfer[]{
  \lsequent{}{\tanh(0)^2 < 1} \quad \lsequent{\tanh(\lambda v)^2 {<} 1}{\dbox{\pchoice{\{\pevolvein{\D{v}=1}{v \leq x}\}}{\{\pevolvein{\D{v}=-1}{v \geq x}\}}}{\tanh(\lambda v)^2 {<} 1}}
  }
  {\lsequent{}{\tanh(\lambda x)^2 < 1}}
\end{align*}}%

The left premise proves using the initial value lemma for $\tanh$, i.e., $\tanh(0)=0$, while the right premise is proved by ODE invariance reasoning with the differential axiom for $\tanh$, i.e., $\der{\tanh(\etermA)} = (1-\tanh(\etermA)^2)\der{\etermA}$~\cite{DBLP:journals/jacm/PlatzerT20}.
Both initial value lemma and differential axiom for $\tanh$ are derived automatically by the definition package (\rref{subsec:impllemma}).

\paragraph{Bouncing ball on sinusoidal wave surface.} The following hybrid program model $\exball$ of a ball bouncing on a sine wave is drawn from the literature~\cite{DBLP:phd/ethos/Denman15,DBLP:conf/fm/0009ZZZ15}.
Compared to the literature, the model $\exball$ given below is parametric in $g > 0$ and proportion $0 \leq k \leq 1$ of the energy lost in an inelastic collision.
  {\small\begin{align*}
  \exball \mnodefequiv \prepeat{\left(\begin{aligned}
  &\{\pevolvein{\D{x} = v_x, \D{y} = v_y, \D{v_x} = 0, \D{v_y} = -g}{y \geq \sin(x)}\} ;\\
  &\keywordfont{if}\left(y = \sin(x)\right)\,
  \left\{\begin{aligned}
     &\pumod{\hat{v}_x}{v_x};\\
     &\pumod{v_x}{\frac{(1-k\cos(x)^2)v_x + (1+k)\cos(x)v_y}{1 + \cos(x)^2}};\\
     &\pumod{v_y}{\frac{(1+k)\cos(x)\hat{v}_x + (\cos(x)^2-k)v_y}{1 + \cos(x)^2}}
  \end{aligned}\right\}
  \end{aligned}\right)}
  \end{align*}}%

Denman~\cite{DBLP:phd/ethos/Denman15} verifies\footnote{Denman~\cite{DBLP:phd/ethos/Denman15} proves the ball height is bounded by $y < 1$ but without bounds on $x$.} a safety property of $\exball$: if the ball is released at rest within a trough $y < 1 \land \frac{\pi}{2} < x < \frac{5\pi}{2}$ of the sinusoidal surface, then it always stays within that trough.
This safety property is specified as follows:
{\small\begin{align*}
\exballfml \mnodefequiv g > 0 \land 0 \leq k \leq 1 \land y < 1 \land \frac{\pi}{2} < x < \frac{5\pi}{2} \land v_x = 0 \land v_y = 0 \limply \dbox{\exball}{\left(y < 1 \land \frac{\pi}{2} < x < \frac{5\pi}{2} \right)}
\end{align*}}%

The \KeYmaeraX verification of $\exballfml$ uses a bound $gy+\frac{1}{2}(v_x^2+v_y^2) < g$ on the total energy of the system, similar to $\exswingfml$.
The main proof step is to show that the kinetic energy is reduced on an inelastic collision (or preserved on a fully elastic collision $k=1$), as modeled by the discrete assignments in $\exball$.
This results in the following (simplified) arithmetic subgoal, which proves automatically using the variable abstraction $c \mnodefequiv \cos(x)$:
{\small\begin{align*}
0 \leq k \leq 1 \limply \left(\frac{(1-kc^2)v_x + (1+k)cv_y}{1 + c^2}\right)^2 + \left(\frac{(1+k)cv_x + (c^2-k)v_y}{1 + c^2}\right)^2 \leq v_x^2 + v_y^2
\end{align*}}

Notably, the example is verified fully parametrically in the constants $g,k$, which makes its direct verification out of reach for numerical techniques~\cite{DBLP:phd/ethos/Denman15,DBLP:conf/fm/0009ZZZ15}.

\paragraph{Robot collision avoidance.}

We model passive orientation safety in robot collision avoidance \cite{DBLP:journals/ijrr/MitschGVP17} to analyze responsibility in collisions with an account of the vision limits of the robot: we consider a robot responsible for collisions with obstacles if it could have stopped before the collision point, or if it ignored the vision limits, but it does not need to actively step out of the way.

\begin{figure}[htb]
  \begin{center}
  \subfloat[Angles along curve of radius $r$, progress $\theta$, angular velocity $\omega$\label{fig:robix-trajectory}]{
  \includegraphics[width=.35\textwidth]{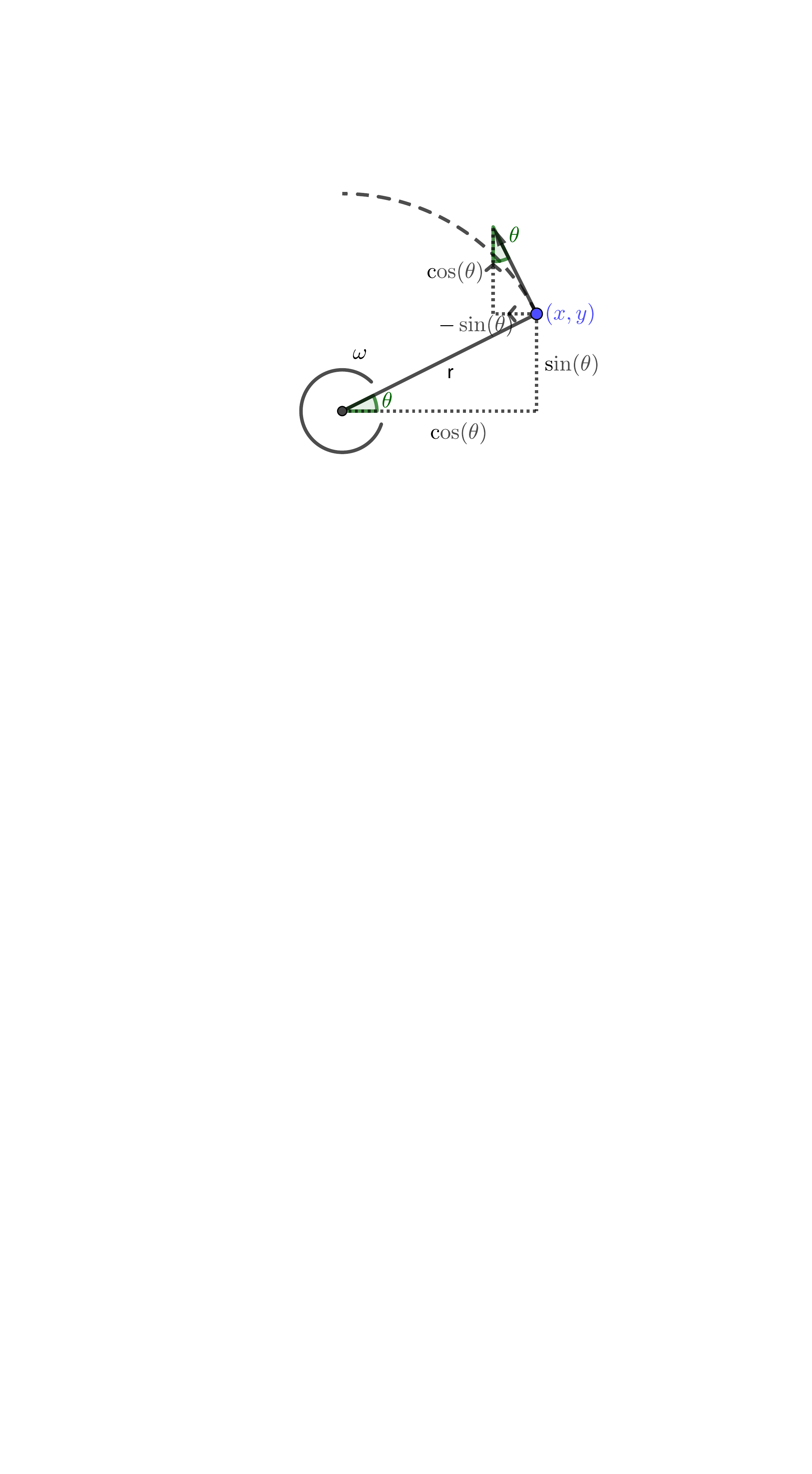}
  } \qquad
  \subfloat[Vision limits of robot, obstacle in distance $d$ is visible when $\sin(\eta) \geq \sin\left(\frac{\pi}{2}-\frac{\gamma}{2}\right)$\label{fig:robix-vision}]{
  \includegraphics[width=.35\textwidth]{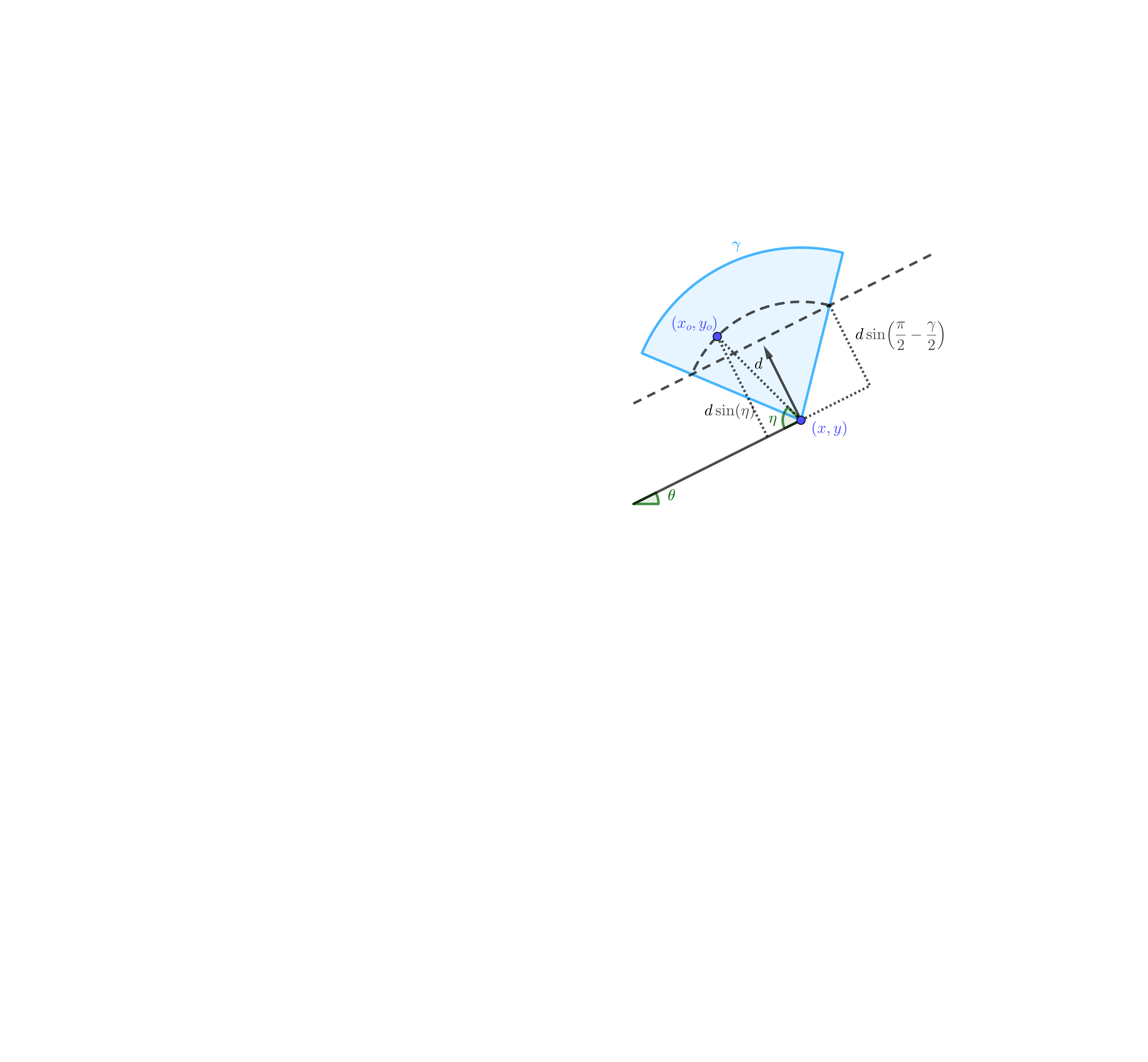}
  }
  \end{center}
\caption{Trajectory and vision limits of a robot at position $(x,y)$ driving along a curve of radius $r$.}
\label{fig:robix}
\end{figure}

The motion of the robot and obstacle are modeled in the differential equations below, where $(x,y)$ is the position of the robot that changes according to the trajectory depicted in \rref{fig:robix-trajectory}, $s \geq 0$ is its driving speed that changes with acceleration $a$, $\theta$ is the angle measuring progress along the trajectory, and $\omega$ the angular velocity influenced by steering $\frac{a}{r}$.
The obstacle is at position $(x_o,y_o)$ and drives in a straight line with velocity vector $(v_x,v_y)$.
  {\small\begin{align*}
    \alpha_r \equiv \D{x}&=-\sin(\theta)s, \D{y}=\cos(\theta)s, \D{s}=a, \D{\theta}=\omega, \D{\omega}=\frac{a}{r}, \D{x}_o=v_x, \D{y}_o=v_y \&s \geq 0
    \end{align*}}

The vision limit angle $\gamma$ of the robot extends symmetrically to the left and right of its direction vector and the obstacle is $d=\sqrt{(x_o-x)^2+(y_o-y)^2}$ distance away from the robot, see \rref{fig:robix-vision}.
In order to determine whether an obstacle is visible to the robot, we compute $\sin(\eta)$ by translating and rotating $(x_o,y_o)$ into the robot's local coordinate frame (robot at $(0,0)$ and direction pointing upwards), and then compare to $\sin\left(\frac{\pi}{2}-\frac{\gamma}{2}\right)$ of the vision limit angle $\gamma$:
{\small\begin{align*}
  \text{visible} \equiv \underbrace{\frac{\sin(\theta)(x_o-x) + \cos(\theta)(y_o-y)}{\sqrt{(x_o-x)^2+(y_o-y)^2}}}_{\sin(\eta)} \geq \sin\left(\frac{\pi}{2}-\frac{\gamma}{2}\right)
\end{align*}}%

The robot controller is allowed to steer and accelerate only if it can stop before hitting any obstacles in its visible range (condition $\text{safe}_d$), and before leaving the area that was visible when making the decision (condition $\text{safe}_r$), otherwise it must use emergency braking:
{\small\begin{align*}
  \text{safe}_d &\equiv \text{visible} \limply {\norm{(x_o,y_o),(x,y)}}_\infty > \overbrace{\frac{v^2}{2b} + \left(\frac{A}{b}{+}1\right)\left(\frac{A}{2}\varepsilon^2 + \varepsilon (v+V)\right) + V\frac{v}{b}}^{\text{collision distance}} \\
  \text{safe}_r &\equiv r\gamma > \underbrace{\frac{v^2}{2b} + \left(\frac{A}{b}+1\right)\left(\frac{A}{2}\varepsilon^2 + \varepsilon v\right)}_{\text{trajectory distance}}
\end{align*}}%

The collision distance and trajectory distance are computed conservatively using maximum acceleration $A$ for the full reaction time $\varepsilon$ followed by full braking $b$ when obstacles approach with maximum speed $V$.
For simplicity, the model is phrased in terms of ${\norm{\cdot}}_\infty$, and so the verification succeeds by exploiting $\sin^2(\theta)+\cos^2(\theta)=1$ and by remembering the state at the time of the robot decision in order to determine who is at fault in case of a collision.
\fi

\end{document}